\documentclass[aps,prb,reprint,floatfix,citeautoscript,longbibliography,superscriptaddress]{revtex4-1}
\usepackage{graphicx,float}
\usepackage{bm,amsmath,amssymb,mathrsfs,dcolumn}
\usepackage{multirow}
\usepackage[usenames]{color}
\usepackage{subfigure}
\usepackage{ulem}
\usepackage{soul,xcolor}
\usepackage{hyperref}
\hypersetup{colorlinks=true, linkcolor=blue, citecolor=blue, urlcolor=blue}
\setstcolor{red}

\begin{document}

\title{Universal field dependence of magnetic resonance near zero frequency}
\author{H. Y. Yuan}
\affiliation{Institute for Theoretical Physics, Utrecht University, 3584 CC Utrecht, The Netherlands}
\author{Rembert A. Duine}
\affiliation{Institute for Theoretical Physics, Utrecht University, 3584 CC Utrecht, The Netherlands}
\affiliation{Center for Quantum Spintronics, Department of Physics, Norwegian University of Science and Technology, NO-7491 Trondheim, Norway}
\affiliation{Department of Applied Physics, Eindhoven University of Technology, P.O. Box 513, 5600 MB Eindhoven, The Netherlands}
\date{\today}

\begin{abstract}
Magnetic resonance is a widely-established phenomenon that probes magnetic properties such as magnetic damping and anisotropy. Even though the typical resonance frequency of a magnet ranges from gigahertz to terahertz, experiments also report the resonance near zero frequency in a large class of magnets. Here we revisit this phenomenon by analyzing the symmetry of the system and find that the resonance frequency ($\omega$) follows a universal power law $\omega \varpropto |H-H_c|^p$, where $H_c$ is the critical field at which the resonance frequency is zero. When the magnet preserves the rotational symmetry around the external field ($H$), $p = 1$. Otherwise, $p=1/2$. The magnon excitations are gapped above $H_c$, gapless at $H_c$ and gapped again below $H_c$. The zero frequency is often accompanied by a reorientation transition in the magnetization. For the case that $p=1/2$, this transition is described by a Landau theory for second-order phase transitions. We further show that the spin current driven by thermal gradient and spin-orbit effects can be significantly enhanced when the resonance frequency is close to zero, which can be measured electrically by converting the spin current into electric signals. This may provide an experimentally accessible way to characterize the critical field. Our findings provide a unified understanding of the magnetization dynamics near the critical field, and may, furthermore, inspire the study of magnon transport near magnetic transitions.
\end{abstract}

\maketitle
\begin{figure}
\centering
\includegraphics[width=0.4\textwidth]{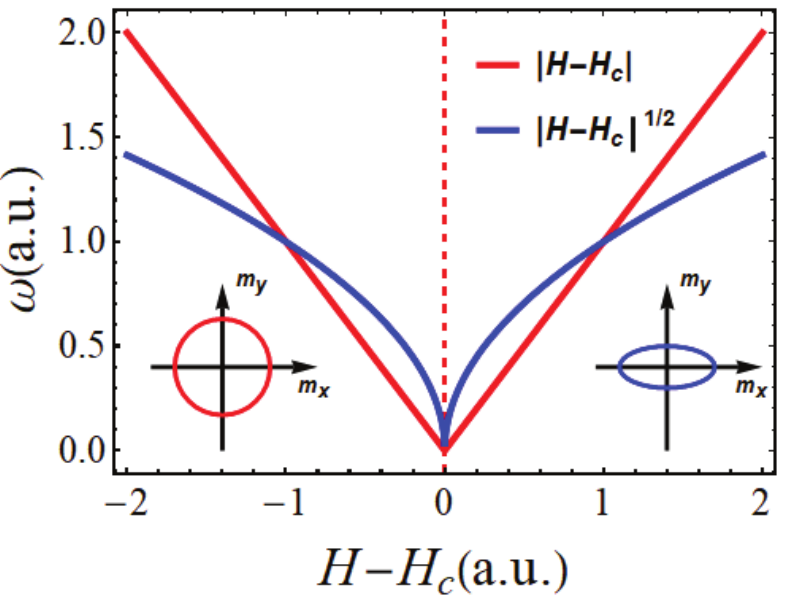}
\caption{Schematic of the magnetic resonance near zero frequency. Depending on the symmetry of the system, the exponent can be either 1 or 1/2.}
\label{fig1}
\end{figure}

{\it Introduction.} Magnetic resonance is the resonant response of a magnetic medium to a driving microwave \cite{Kittel1948} and it is an important knob to manipulate the spin dynamics and spin transport. The resonance frequency of a ferromagnet (FM) is determined by the external field and anisotropy of the system, and is usually on the order of a few gigahertz. In an antiferromagnet (AFM), the resonance frequency can reach the terahertz regime due to the strong exchange coupling between the two or more magnetic sublattices. On the other hand, experiments have observed that the resonant signal can range from megahertz down to the kilohertz regime \cite{Hagiwara1999,Manuilov2010,Qin2018,Chavez2013,Hils2013}, and it may be lowered further close to zero resulting in the excitation of a soft spin mode (as shown in Fig. \ref{fig1}). This usually happens when the magnetization reorients itself to find new equilibrium states \cite{Mont2008,Zhang2017}. For example, the antiferromagnetic spin flops from an easy-axis to a hard-plane when the applied field is strong enough and this gives rise to the zero resonance frequency at the transition point \cite{Keffer1952}. According to the modern theory of phase transitions \cite{Landau}, the change of the ground state of a system in the thermodynamic limit may be accompanied by critical phenomena and there exists a set of critical exponents characterizing the scaling behavior of the physical quantities near the transition point. A well-known example is the ferromagnetic phase transition from the paramagnetic state to a ferromagnetic state at the Curie temperature,
at which the scaling law of the spin wave dispersion versus temperature has been well studied \cite{Halperin1969}. However, it is not clear whether the magnetic resonance frequency generally follows a universal power law near zero frequency. Some pioneering works in both theory and experiments have addressed this issue in special geometries including elliptical nanodots \cite{Mont2008}, magnetic nanostrip \cite{Bail2003,Zeng2016} and magnetic monolayer \cite{Pini2005}, while a unified picture is lacking.

In this article, we derive a generalized power law of the magnetic resonance near the zero resonance frequency, i.e. $\omega \varpropto |H-H_c|^p$, where $H$ is the external applied field and $H_c$ is the critical field. Here the exponent $p$ can be fully determined by the symmetry properties of the magnetic system, and it is independent of values of magnetic parameters. When the magnetization reorients itself continuously near the critical field, a second-order phase transition is identified, which is well described by the Landau theory for second-order phase transitions. In samples in the thermodynamic limit, and at finite temperature or when quantum fluctuations are important, we therefore expect the exponent $p$ becomes $z\nu$, with $z$ the dynamical critical exponent and $\nu$ the exponent that characterizes the divergence of the correlation length \cite{Hoh1977}. Furthermore, such a phase transition can enhance the spin current transport at cryogenic conditions driven by either thermal gradient or spin-orbit effects and thus can be measured electrically by transforming the spin current into charge current. These findings may provide a unified picture to understand the transition behaviors in magnetic resonance, and may further benefit its usage in studying magnon transport properties.

{\it Universal power law.} We consider a magnet with equilibrium magnetization direction ($\mathbf{m}$) along the $z$ axis. The magnetization dynamics under perturbations is described by the Landau-Lifshitz-Gilbert (LLG) equation,
\begin{equation}
\frac{\partial \mathbf{m}}{\partial t}=-\gamma \mathbf{m} \times \mathbf{H}_{\mathrm{eff}} + \alpha \mathbf{m} \times \frac{\partial \mathbf{m}}{\partial t},
\end{equation}
where $\gamma >0$ is the gyromagnetic ratio, $\mathbf{H}_{\mathrm{eff}}=-\delta \mathcal{H}/\delta \mathbf{m}$ is the effective field acting on the magnetization with $\mathcal{H}$ being the Hamiltonian of the system, and $\alpha$ is the Gilbert damping. To obtain the resonance frequency, we consider a small fluctuation around the equilibrium magnetization, i.e. $\mathbf{m}=e_z + (m_xe_x + m_y e_y)e^{-i\omega t}$ and linearize the LLG equation as,
\begin{equation}
\left ( \begin{array}{cc}
  i\omega - \gamma E_{xy} & -\gamma E_{yy}+i\alpha\omega \\
  \gamma E_{xx} -i\alpha\omega& i\omega +\gamma E_{xy}
\end{array} \right )
\left (\begin{array}{c}
  m_x \\
  m_y
\end{array} \right )=0,
\label{secular}
\end{equation}
where $E_{ij}= \partial \mathcal{H}/\partial m_i \partial m_j$. Around the equilibrium magnetizaiton, the total energy can be expanded as $\mathcal{H}=\mathcal{H}_0+1/2 \sum E_{ij}m_im_j$, where $\mathcal{H}_0$ is the ground state energy, and the first-order derivatives  vanish ($\partial \mathcal{H}/\partial m_i=0$) to minimize the total energy. By solving the secular equation (\ref{secular}), we derive the resonance frequency in the absence of damping as \cite{Smit1955},
\begin{equation}
\omega/\gamma = \sqrt{E_{xx}E_{yy}-E_{xy}^2}.
\end{equation}
Here we shall focus on the resonance behavior close to zero frequency, i.e. $\omega \sim 0$.

If the magnetic system obeys rotational symmetry around the equilibrium magnetization ($e_z$), then $\mathcal{H}=\mathcal{H}(m_z)=\mathcal{H}(1-(m_x^2+m_y^2)/2)$. This implies that the expansion of $\mathcal{H}$ with respect to $m_x$ and $m_y$ should be symmetric and have no crossing terms $E_{xy}$, i.e. $E_{xy}=0, E_{xx}=E_{yy}$. A further Taylor expansion of $E_{ii}$ at the critical field suggests that the leading-order contribution is $E_{xx}=E_{yy}\propto (H-H_c)$. Then we immediately see that $\omega \propto |H-H_c|$. One example is the resonance of an uniform magnetic sphere under the external field. From the Hamiltonian $\mathcal{H}=-m_z H$, it is straightforward to find $E_{xx}=E_{yy}=H$ and thus $\omega/\gamma = |H|$.

If the system breaks the rotational symmetry with respect to the equilibrium magnetization, the anisotropy energy will generate non-zero $E_{xx}$, $E_{yy}$ and $E_{xy}$, in principle. However, we can always choose two principal axis ($x'y'$) by rotating the $xy$ coordinates to eliminate the term $E_{xy}$. Now the energy landscape in the $x'$ and $y'$ directions will differ. When $E_{x'x'}$ ($E_{y'y'}$) approaches zero by tuning the external field, $E_{y'y'}$ ($E_{x'x'}$) is not zero. Therefore the resonance frequency will depend on the field as as $\omega \propto |H-H_c|^{1/2}$.

Table \ref{tab1} summarizes the resonance frequency in the commonly used magnetic systems without and with crystalline anisotropy. All of them are well described by the result $\omega \propto |H-H_c|^{p}$ with a symmetry dependent exponent $p$.

\begin{widetext}
\begin{table*}
\centering
\caption{Summary of the resonance frequency and critical exponent in the magnetic systems without (top panel) and with (bottom panel) crystalline anisotropy. $H_E$ is the exchange field and $H_{\mathrm{sp}}$ is spin-flop field for antiferromagnets. Y and N refers to systems which have and do not have rotational symmetry around the equilibrium magnetization.}
\begin{tabular}{l|c|c|c|c|c|c}
    \hline
    \hline
  Anisotropy & Direction &Field $H$ & Resonance frequency $\omega/\gamma$ & Critical field $H_c$& Exponent $p$ &Symmetry\\
  \hline
    Sphere &Isotropic& Arbitrary &  $|H|$ & 0 & 1 & Y\\
    \hline
  Thin & Easy&Normal & $H-M_s$ & $M_s$ & 1 & Y\\
   disk&plane& In-plane & $\sqrt{H(H+M_s)}$ & 0 & 1/2 &N\\
    \hline
  Long  &Easy& Axial &$H+M_s/2$ & $-M_s/2$ & 1 & Y\\
   cylinder&axis& Radial& $\sqrt{H(H-M_s/2)}$ & $M_s/2$ & 1/2 & N\\
  \hline
  \hline
  \hline
  \multirow{2}{*}{Uniaxial}  &\multirow{2}{*}{Easy $z$} & $z$ &  $H+2K_z$ & 0 & 1 &Y\\ \cline{3-7}

                             & &$x$ &
  $\begin{array}{ll}
                          \sqrt{(2K-H)(2K+H)}, & H<H_c \\
                          \sqrt{H(H-2K)}, & H\geq H_c
                        \end{array}$
    &$2K$& 1/2 & N \\
    \hline
  \multirow{2}{*}{Biaxial} & Easy $z$&$z$ & $\sqrt{(H+2K_z+2K_x)(H+2K_z)}$ & $-2K_z$ & 1/2 &N\\ \cline{3-7}
   & Hard $x$ &$x$ & $\begin{array}{ll}
                          \sqrt{2K_z(H_c-H^2/H_c)}, & H<H_c \\
                          \sqrt{(H-H_c)(H-2K_x)}, & H\geq H_c
                        \end{array}$ & $2(K_z+K_x)$ & 1/2 &N\\ \cline{3-7}
    \hline
  \multirow{2}{*}{Cubic} & &(111) &$H-4K_1/3$ & $4K_1/3$ & 1 &Y\\ \cline{3-7}
  & &(001) &$H+2K_1$ & $-2K_1$ & 1 &Y\\ \cline{3-7}
  & &(110) &$\sqrt{(H+K_1)(H-2K_1)}$ & $2K_1$ & 1/2 &N\\ \cline{3-7}
  \hline
\multirow{2}{*}{\begin{tabular}{c}
                                  Uniaxial \\
                                  AFM \\
                                \end{tabular}}
   & \multirow{2}{*}{Easy $z$} &  $z$ & $\begin{array}{ll}
                          -H+H_{\mathrm{sp}}, & H<H_c \\
                          \sqrt{H^2-H_{\mathrm{sp}}^2}, & H\geq H_c
                        \end{array}$ & $H_{\mathrm{sp}}$ & $\begin{array}{ll}
                         1, & H<H_c \\
                          1/2, & H \geq H_c
                        \end{array}$ &\begin{tabular}{c}
                                  Y \\
                                  N \\
                                \end{tabular}\\ \cline{3-7}
   &  &  $x$ & $\begin{array}{ll}
                          \sqrt{H_{\mathrm{sp}}^2-2K_z/H_cH^2}, & H<H_c \\
                          \sqrt{(H-H_c)(H-H_c/2)}, & H\geq H_c
                        \end{array}$ & $2H_{E}$ & 1/2 &\begin{tabular}{c}
                                  N \\
                                  N \\
                                \end{tabular}\\ \cline{3-6}
  \hline
  \begin{tabular}{c}
                                  Uniaxial \\
                                  +Demag$\cite{Smit1955}$\\
                                \end{tabular}
   & Easy $z$ &  $x$ & $\sqrt{(2K+M_s)H_c(1-H^2/H_c^2)}$ & $2K+M_s/2$ & 1/2&N\\ \cline{3-5}
  \hline
\end{tabular}
\label{tab1}
\end{table*}
\end{widetext}

{\it Phase transition.} The behavior of magnetization near the critical field may be interpreted as a characterization of phase transition. When $p=1$ below or above the critical field, the transition is first-order, because it is accompanied by a sudden switching of the magnetic order. For example, in the uniaxial case (so-called Stoner-Wohlfarth model), a field along the easy axis ($e_z$) will induce the magnetic switching below the transition point $H_c=-2K_z$.

\begin{figure}
\centering
\includegraphics[width=0.45\textwidth]{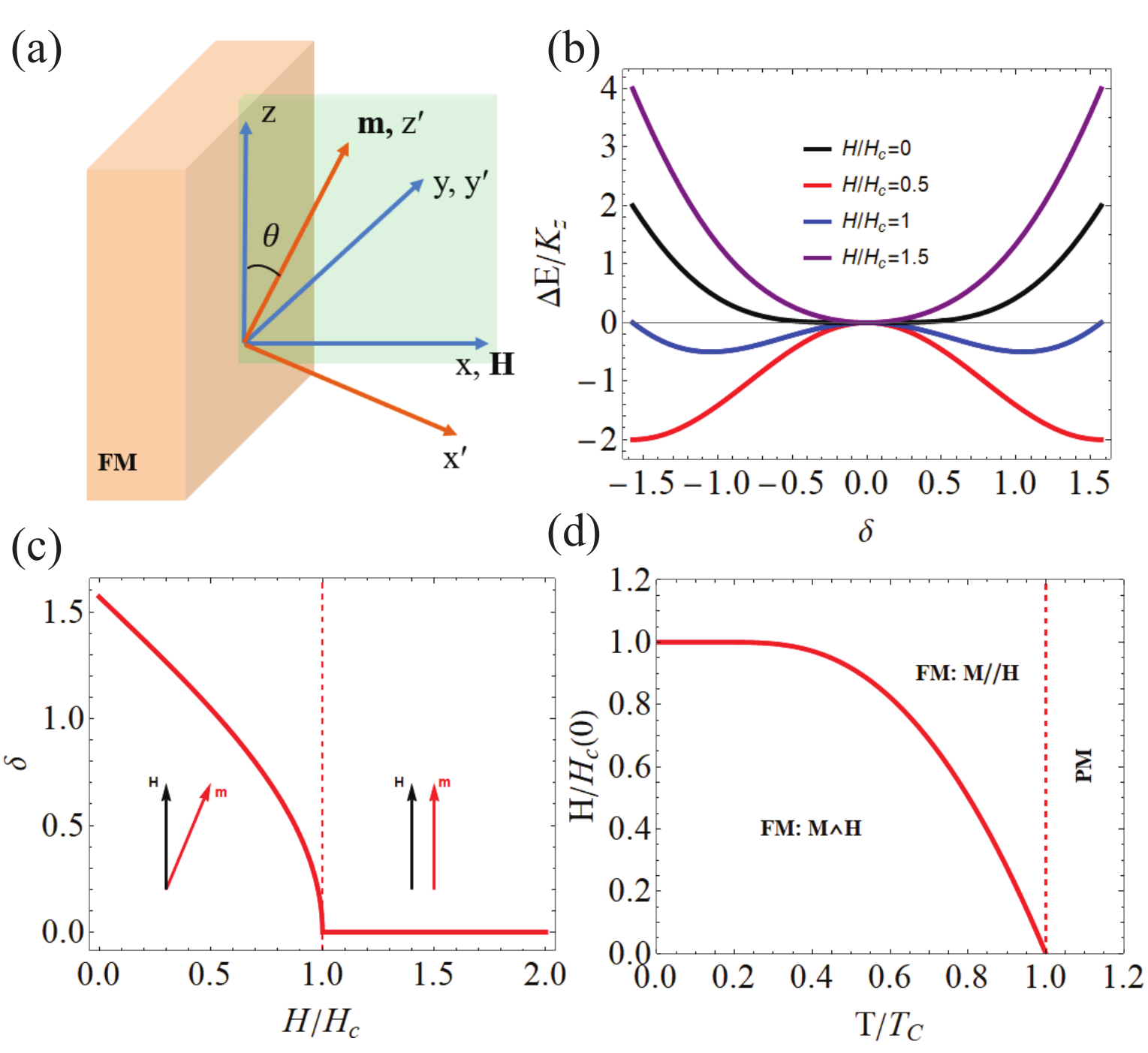}
\caption{(a) Schematic of the biaxial magnetic system with an easy axis $x$ and a hard axis $z$. (b) Energy landscape as a function of the order parameter $\delta$. (c) Order parameter $\delta$ as a function of external field.  (d) Phase diagram of a biaxial-magnet in the ($T/T_c, H/H_c(0)$) plane. FM: ferromagnet, PM: paramagnet.}
\label{fig2}
\end{figure}

When $p=1/2$ both below and above the critical field, the magnetization reorients continuously as the field is tuned across the transition point, it will correspond to a second-order phase transition. This point is justified by expanding the total energy around the transition point as,
\begin{equation}
\mathcal{H} = \mathcal{H}_0 +a_2 \delta m^2 + a_4 \delta m^4 +O(\delta m^6),
\end{equation}
where $\delta m$ is a small deviation from the equilibrium direction. Here $a_2\propto E_{x'x'} \propto H-H_c$ will change sign at the transition field, the $\delta m^3$ term is absent because the anisotropy energy of a magnet is an even function of magnetization, $a_4 >0$ to guarantee that the system has a stable magnetization when $H<H_c$. One immediately sees that the behavior of this Hamiltonian can be well described by the Landau theory of the second-order phase transition \cite{Landau}. The explicit form of the expansion coefficients for an uniaxial ferromagnet was first discussed by Wang \textit{et al.}\cite{Zeng2016}. To illustrate this point, let us further take a biaxial magnet as well as an antiferromagnet as examples. The energy of a biaxial system shown in Fig. \ref{fig2}(a) is,
\begin{equation}
\begin{aligned}
\mathcal{H} =& -K_z \cos^2\theta + K_x \sin^2 \theta \\
&- H (\cos\theta \cos \theta_H +\sin\theta \sin \theta_H \cos (\varphi-\varphi_H)),
\end{aligned}
\end{equation}
where $K_z,K_x>0$ are respectively the easy and hard axis anisotropy, $\theta_H$ and $\varphi_H$ are the polar and azimuthal angles of the external field. By taking $\partial \mathcal{H}/\partial \theta=0,\partial \mathcal{H}/\partial \varphi=0$, the equilibrium magnetization should satisfy $\varphi_0=\varphi_H$ and
\begin{equation}
(K_z+K_x)\sin 2 \theta_0 =  H \sin(\theta_H - \theta_0 ).
\end{equation}
When the field is along the hard-axis $\theta_H = \pi/2,\varphi_H = 0$, the ground state is,
\begin{equation}
\theta_0=  \left \{ \begin{array}{cc}
            \arcsin H/H_c, & H<H_c, \\
            \pi/2, & H \ge H_c,
          \end{array} \right.
          \label{equil}
\end{equation}
where the critical field $H_c =2(K_z+K_x)$. To clarify the transition behavior near $H_c$, we reformulate the Hamiltonian as a polynomial of the order parameter as,
\begin{equation}
\mathcal{H}=K_x-H+\frac{1}{2}(H-H_c) \delta^2+ \frac{1}{24}(4H_c-H)\delta^4,
\end{equation}
where the magnetic orientation $\delta =\theta-\pi/2$ may be viewed as the order parameter of the system.
Figure \ref{fig2}(b) shows the energy profile as a function of $\delta$. Below $H_c$, there are two energy minima of the system ($\theta_0$ and $\pi-\theta_0$). Since the external field breaks the $\mathcal{Z}_2$ symmetry of the system ($m_x\rightarrow-m_x,m_y\rightarrow-m_y$), the magnet will stabilize at one particular minimum ($\theta_0$ for $H>0$). As the external field increases above $H_c$, the two minima merge into a single minimum at $\theta_0=\pi/2$ where the $\mathcal{Z}_2$ symmetry is recovered, as shown in Fig. \ref{fig2}(c). This is a clear feature of second-order phase transition associated with $\mathcal{Z}_2$ symmetry breaking, but the driving force is an external field instead of temperature. Note that a simplified version of this formalism ($K_x=0$) resembles the mean field approach of the disorder-order phase transition in a transverse Ising model \cite{Stin1973}. The difference is that here we have the transition between the two ordered phases, while the quantum fluctuation generates a disordered phase in the Ising model.

When non-zero temperatures are considered, both the anisotropy strength ($K$) and saturation magnetization ($M_s$) will decrease with temperature. Considering the well-known power law $K\propto m^3$, $M_s \propto m$ \cite{Callen1966}, the critical field scales as $H_c \propto m^2$, where $m=M_s(T)/M_s(T=0)$ is a normalized magnetization. Here $m$ can be obtained from the mean field theory by solving the equation $m=\tanh(mT_c/T)$ in a self-consistent way, where $T_c$ is the Curie temperature. Figure \ref{fig2}(d) shows the phase diagram of the system in ($T/T_c, H/H_c(T=0)$) plane. As the temperature increases above $T_c$, the system has a second-order phase transition to become a paramagnetic state. Below the Curie temperature, there is a continuous phase boundary (red line) which separates the system into two phases according to $\mathcal{Z}_2$ symmetry of their ground states. As a comparison, for an easy-plane magnet ($K_z=0,K_x\neq 0$), the corresponding phase diagram is similar. However, we will have a $U(1)$ symmetry breaking below the phase boundary, where the resulting dynamics of the in-plane magnetization gives rise to
a spin superfluidity \cite{Flebus2016}.

Another example is the two-sublattice uniaxial AFM subject to a transverse field, the Hamiltonian is $\mathcal{H}=H_E \mathbf{m}_1 \cdot \mathbf{m}_2 -K_z(m_{1z}^2 + m_{2z}^2)-H(m_{1x} + m_{2x})$, where $H_E$ is exchange field,
$\mathbf{m}_1$ and $\mathbf{m}_2$ are the magnetization on the two sublattices. Close to the critical field $H_c=2H_E$ (as indicated in Table \ref{tab1}), the Hamiltonian can be expanded as,
$\mathcal{H}=-2H+H_E + (H-H_c)\delta^2+(4H_E-H)/12\delta^4$, which is similar to the biaxial case, and thus also hosts a second-order phase transition. Recently, this transition was observed in an antiferromagnetic $\mathrm{CrI_3}$ bilayer \cite{Zhang2020}.

\begin{figure}
\centering
\includegraphics[width=0.45\textwidth]{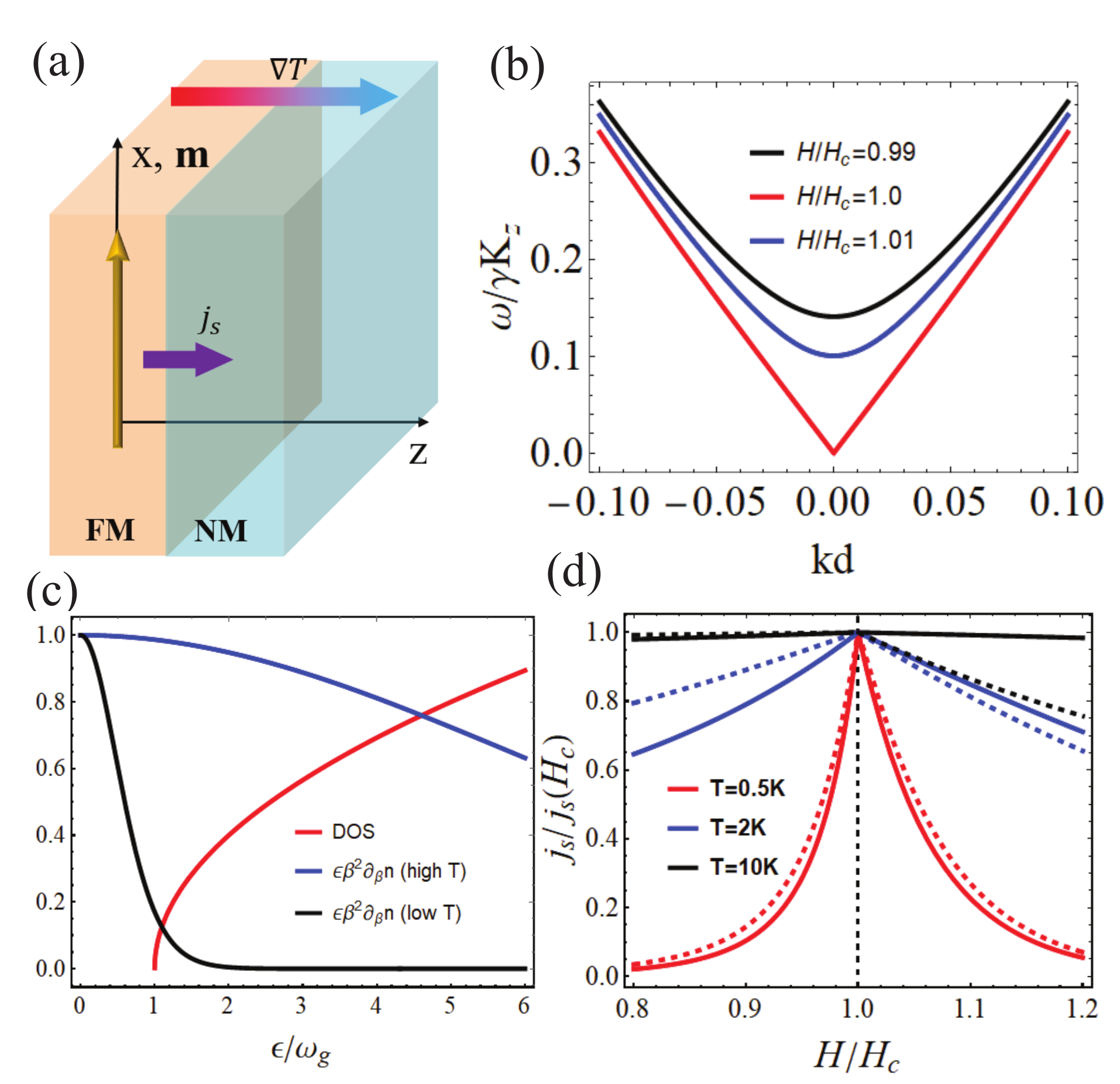}
\caption{(a) Schematic of the spin Seebeck setup in the FM/NM bilayer. (b) Spin wave spectrum near the transition point. $K_x=0$.
(c) Integrand of Eq. (\ref{sse}) as a function of magnon energy. (d) Thermal spin current as a function of external field at $T=$ 0.5 K (red line), 2 K (blue line), and 10 K (black line), respectively. The curved dashed lines are calculated based on analytical formula (\ref{sse_th}).}
\label{fig3}
\end{figure}

{\it Spin injection by spin Seebeck effect.}
Modern spintronics, among other subjects, concerns the transport properties of magnons for information processing \cite{Bauer2012,Chumak2015}. Hence it may be meaningful to study whether the magnon spin current can be manipulated at the transition point. To address this issue, we consider a ferromagnet$|$normal magnet (FM$|$NM) bilayer subject to a thermal gradient as sketched in Fig. \ref{fig3}(a). The Hamiltonian of the magnetic layer is,
\begin{equation}
\mathcal{H}=-J\sum_{\langle i,j\rangle} \mathbf{S}_i\cdot \mathbf{S}_j + K_x \sum_j (S_j^x)^2-\sum_j K_z(S_j^z)^2-H\sum_j S_j^x,
\label{hamsse}
\end{equation}
where $J$ is the exchange stiffness, $\mathbf{S}_j$ is the spin vector at the $j-$th site with magnitude $|\mathbf{S}_j|=S$ and the first sum is taken over all the nearest neighbors.
We shall first solve the magnon spectrum of the system and then apply it to calculate the thermal spin current.

The magnon spectrum can be obtained by first transforming the magnetic system to a new frame, with $z'$ polarized at the equilibrium $\mathbf{m}$ and $y'=y$, through the rotational operation,
\begin{equation}
\left (\begin{array}{c}
  S_{j,x} \\
  S_{j,y} \\
  S_{j,z}
\end{array} \right )=
\left ( \begin{array}{ccc}
  \cos \theta & 0 & \sin \theta \\
  0 & 1 & 0 \\
  -\sin \theta & 0  &\cos \theta
\end{array} \right )
\left ( \begin{array}{c}
  S_{j,x'} \\
  S_{j,y'} \\
  S_{j,z'}
\end{array} \right ),
\label{rotation}
\end{equation}
where $S_{j,z'}=S-a_j^\dagger a_j, S_{j,x'}=\sqrt{2S}(a_j + a^\dagger_j)/2,S_{j,y'}=\sqrt{2S}(a_j-a_j^\dagger)/2i$ \cite{HP1940}. $a_j$ ($a_j^\dagger$) is the magnon annihilation (creation) operator on the $j-$th spin site. By substituing
Eq. (\ref{rotation}) into the Eq. (\ref{sse}) and further transferring to the momentum space, the Hamiltonian (\ref{sse}) can be recast as,
\begin{equation}
\mathcal{H}=\sum_{k>0} \left (\omega_{a,k} a_k^\dagger a_k + \omega_{a,-k} a_{-k}^\dagger a_{-k} + g(a_k^\dagger a_{-k}^\dagger + a_k a_{-k}) \right ),
\label{qmH}
\end{equation}
where $\omega_{a,k} =\omega_{a,-k}= 2ZJS(1-\gamma_k) + H \cos \theta_0 + 2K_zS \sin^2\theta_0 -K_zS \cos^2 \theta_0,g=(K_x-K_z)S\cos^2\theta_0$, $Z$ is coordinate number, $J$ is the exchange strength, $\gamma_k=1/z \sum e^{i\mathbf{k}\cdot \mathbf{d}}$, where the sum is taken over all the nearest neighboring cells and $d =|\mathbf{d}|$ is the lattice constant.
By diagonalising the Hamiltonian (\ref{qmH}) using a Bogoliubov transformation, similar to the spectrum calculation of antiferromagnetic magnons \cite{yuan2020}, the low energy magnon spectrum is found as,
\begin{equation}
\omega=\sqrt{(ZJSd^2 k^2 +H-2K_xS)(ZJSd^2 k^2+(H-H_c))},
\end{equation}
for $H>H_c$, and
\begin{equation}
\omega=\sqrt{(ZJSd^2 k^2 +2K_zS)\left(ZJSd^2 k^2+\frac{H_c^2-H^2}{H_c} \right )},
\end{equation}
for $H<H_c$, where the critical field is $H_c=2(K_z+K_x)S$. One immediately sees that the magnon excitation is gapless at the transition point ($H=H_c$) and gapped both above and below $H_c$, as shown in Fig. \ref{fig3}(b).

The thermal spin current across the interface of FM$|$NM shown in Fig. \ref{fig3}(a) is related to the magnon excitation inside the FM layer as \cite{Bender2012,Hoffman2013},
\begin{equation}
j_s=-4k_B \Delta T \alpha' \int_{\epsilon_g}^\infty d\epsilon D(\epsilon)\epsilon \beta^2\frac{\partial n(\beta \epsilon)}{\partial \beta},
\label{sse}
\end{equation}
where $k_B$ is the Boltzman constant, $\alpha'$ is reduced spin-mixing conductance,  $\epsilon=\hbar \omega$, $\hbar$ is Planck constant, $\Delta T$ is the temperature drop at the FM$|$NM interface,
$D(\epsilon)$ is the density of states (DOS) of magnons, $\beta=1/k_BT$, $n(x)=1/(e^{\beta x}-1)$ is the Bose-Einstein distribution. The integration range goes from the band bottom $\epsilon_g$ up to infinity. Note that magnon DOS $D(\epsilon)$ is a monotonically increasing function of energy $\epsilon$ while the remaining integrand $\epsilon\beta^2 \partial n(\beta \epsilon)/\partial \beta$ rapidly decreases to zero as $\epsilon$ increases from 0 by $k_BT$, as shown in Fig. \ref{fig3}(c). Therefore, to make the magnon occupation near the gap $\epsilon_g$ contribute significantly to the spin current, one has to enter into the low temperature regime. On the other hand, the spin current is expected to be enhanced at the transition point because the gapless magnons are more easily to be excited and contribute to the spin current.

By performing the integral (\ref{sse}) numerically, we obtain the spin current as a function of external field as shown in Fig. \ref{fig3}(d). The parameters used are MnGa with $K_z=1.2\times 10^6~\mathrm{J/m^3},M_s=2.5\times 10^5 ~\mathrm{A/m}, T_c=800 ~\mathrm{K}, zJ\approx k_BT_c/(\hbar \gamma)=595 ~\mathrm{T}$ \cite{Mizukami2011}. Clearly, the spin current is maximal at the transition point, and this enhancement increases as the temperature decreases, which are both consistent with the expectations. In the low temperature regime, we may also get analytical results of the spin current by
noticing that (i) mainly long wavelength magnons ($kd\rightarrow 0$) are thermally excited, i.e., the spectrum can be approximated as $ \epsilon=\sqrt{Ak^2 + \epsilon_g^2}$ and (ii) the integrand $\partial n/\partial \beta \approx -\epsilon e^{-\beta \epsilon}$,
where the effective spin wave stiffness is,
\begin{equation}
A=  \left \{ \begin{array}{cc}
            2K_zZJ\hbar^2S^2d^2, & H<H_c, \\
            (H-2K_xS)ZJ\hbar^2 S^2 d^2, & H \ge H_c,
          \end{array} \right.
          \label{spectrum}
\end{equation}
and spin wave gap is,
\begin{equation}
\epsilon_g =  \left \{ \begin{array}{cc}
            \hbar\sqrt{2K_zS(H_c-H)(H_c+H)/H_c}, & H<H_c, \\
            \hbar \sqrt{(H-2K_xS)(H-H_c)}, & H \ge H_c.
          \end{array} \right.
          \label{spectrum}
\end{equation}
Now the DOS of magnons can be derived as $D(\epsilon)=4\pi/A^{3/2} \epsilon \sqrt{\epsilon^2-\epsilon_g^2}$, while the integral in Eq. (\ref{sse}) is analytically evaluated as,
\begin{equation}
j_s=\frac{16\pi k_B \Delta T \alpha' \epsilon_g^2}{\beta A^{3/2}} \left ( 3 \beta \epsilon_g K_1(\beta \epsilon_g) +
(12+\beta^2\epsilon_g^2)K_2 (\beta \epsilon_g)\right ),
\label{sse_th}
\end{equation}
where $K_\nu (x)$ is the modified Bessel functions of the second kind, which decays monotonically with the increase of $x$. Close to the transition point ($\epsilon_g =0$), we can further expand $K_\nu(x)$ and approximate the spin current (\ref{sse_th}) as,
\begin{equation}
j_s=16\pi k_B \Delta T \alpha' \frac{24-\beta^2 \epsilon_g^2}{\beta^3A^{3/2}}.
\end{equation}
Again, we can see that the spin current is maximally excited at the transition point, and it follows a universal field dependence near the transition as $(j_s(H)-j_s(H_c))/j_s(H_c) \approx |H-H_c|$. Note that these analytical results only
work quantitatively well at low temperature as indicated in the comparisons of analytical and numerical results in Fig. \ref{fig3}(d).

\begin{figure}
\centering
\includegraphics[width=0.45\textwidth]{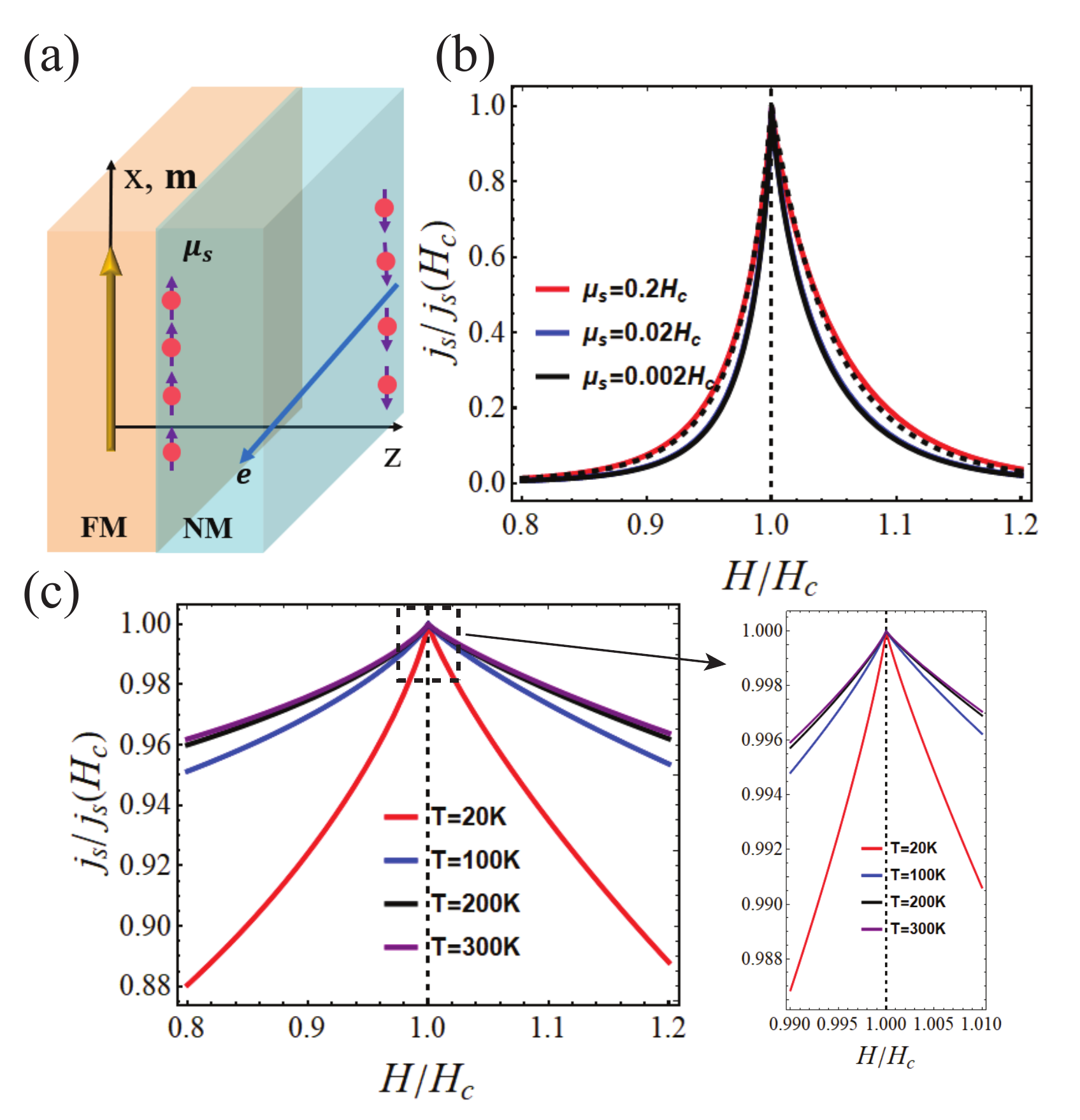}
\caption{(a) Schematic of the spin injection through spin Hall effect. (b) Injected spin current as a function of external field at $\mu_s=0.2H_c$ (red line), $0.02H_c$ (blue line), and $0.002H_c$ (black line), respectively, with a fixed temperature $T=0.5$ K. The curved dashed line is calculated based on the analytical formula (\ref{she_th}). All the other parameters are the same as Fig. \ref{fig3}. (c) Injected spin current as a function of external field at $T=20$ K (red line), 100 K (blue line), 200 K (black line), and 300 K (purple line), respectively, with a fixed spin accumulation $\mu_s=0.02H_c$.}
\label{fig4}
\end{figure}

{\it Spin injection by spin-orbit effects.} Besides applying a thermal gradient, one may also inject spin current through the spin-orbit effects. For example, one can pass electric current in the normal metal layer of FM$|$NM bilayer, which may produce a transverse spin current in the NM layer through spin Hall effect. Such a spin current further generates a spin accumulation ($\mu_s$) at the interface and thus injects spin spin currents into the FM layer (see Fig. \ref{fig4}(a)). The magnitude of the spin current can be formulated as \cite{Bender2012},
\begin{equation}
j_s=4 \alpha' \int_{\epsilon_g}^\infty d\epsilon D(\epsilon)(\epsilon-\mu_s) (n(\beta \epsilon) - n(\beta (\epsilon-\mu_s))).
\label{she}
\end{equation}
Similar to the spin Seebeck setup, the injected spin current is also enhanced near the transition point, as shown in Fig. \ref{fig4}(b). Here the magnitude of enhancement converged as $\mu_s$ reduces. This is because the integrand in Eq. (\ref{she}) can be expanded as $(\epsilon-\mu_s) (n(\beta \epsilon) - n(\beta (\epsilon-\mu_s)))\approx \beta \mu_s\partial n/\partial \beta$ when $\mu_s \rightarrow 0$ and thus the factor $\mu_s$ is scaled out in calculating $j_s/j_s(H_c)$. Analytically, we can derive the spin current in this regime as,
\begin{equation}
j_s = -\frac{16\pi \alpha' \mu_s \epsilon_g^2}{A^{3/2} \beta}\left (  \beta \epsilon_g K_1(\beta \epsilon_g) +3 K_2(\beta \epsilon_g)\right ).
\label{she_th}
\end{equation}
Near the transition point, the spin current can be further approximated as,
\begin{equation}
j_s = -\frac{8\pi \alpha' \mu_s }{A^{3/2} \beta^3}\left (  12- (\beta \epsilon_g)^2\right ).
\end{equation}
Here, the spin current follows a similar power law near transition point as spin seebeck case as $(j_s(H)-j_s(H_c))/j_s(H_c) \approx |H-H_c|$. Note that the absolute value of spin current injected into the FM layer will be very small when spin accumulation at the interface is small because $j_s \propto \mu_s$, which will make it difficult to be measured.  Moreover, for a given spin accumulation, the enhancement decreases with the increase of temperature, as shown in Fig. \ref{fig4}(c). Therefore, a moderate value of spin accumulation at low temperature is preferred to observe this enhancement in experiments.

{\it Discussions and Conclusion.}
In conclusion, we have derived a universal power law of the magnetic resonance frequency near zero resonance frequency, and the exponents depends on the symmetry of the system. When the magnetization reorients continuously around the transition point, it can be viewed as a second-order phase transition. This transition can significantly enhance the spin current injected by a thermal gradient or spin-orbit effects and thus be measured in an electrical way by transforming the spin current into charge current through inverse spin Hall effect or spin Hall magnetoresistence.

Our result is in principle valid for a finite system described as a classical. In the thermodynamic limit, the exponent 1/2 will become $z \nu$. On the other hand, even though we derive the power law based on the macrospin approach, the multi-domain magnetic structure may form at nearly zero resonance frequency. Suppose there are only two types of domains inside the magnet, this power law still works. For example, in a single crystal of hexagonal $\mathrm{BaFe_{12}O_{10}}$ with the coexistence of uniaxial and shape anisotropy, there are two antiparallel domains separated by domain walls \cite{Smit1955}. Since the rotational symmetry is broken in such a structure, the critical exponent is 1/2, as summarized in Table \ref{tab1}. Another similar example is the uniaxial antiferromagnet, it preserves (breaks) the rotational symmetry below (above) the spin flop transition field ($H_\mathrm{sp}$) and thus acquires the critical exponent $p=1 (1/2)$, respectively. When the number of domains increases above two, it has been shown that the critical exponent is 1/2 for an in-plane magnetized multi-domain state \cite{Zeng2016}, but the validity of the power law in the general case with multi-domains separated by various types of domain walls, such as transverse walls and vortex walls, is yet to be studied.

Furthermore, the field-dependence of the magnon frequency can go beyond the magnetic resonance and apply to a wide class of quantum optical systems that involves the effective spin degree of freedom. For example, the Dicke model describes the interaction of a collection of two-level atoms and electromagnetic wave as $\mathcal{H}=\omega_a a^\dagger a +\omega_m S_z +g_{\mathrm{am}} (a+a^\dagger)(S^+ + S^-)$, where $a$ is the photon field and $S^+(S^-)$ is the spin rising (lowering) operator defining as the sum of spin operators on each atom, $\omega_a$, $\omega_m$ and $g_{\mathrm{am}}$ are respectively the frequencies of photon, magnon, and their coupling strength \cite{Dicke1954,Hepp1973,Emary2003}. The frequency of lower eigenmode will approach zero as $\omega \propto \sqrt{\omega_m-\omega_c}$ at the critical frequency $\omega_c=4g_{\mathrm{am}}^2/\omega_a$, which can be understood within our picture. Here the photon field plays a similar role as a transverse field in the uniaxial model and thus it can induce a continuous transition that follows the power law of $p=1/2$.

Lastly, spin current transport near the AFM spin-flop transition has already been reported \cite{Lebrun2018,Bender2017,Li2020,Reitz2020,yuanreview}. All these works show the enhancement of spin current transport near the transition point, which are consistent with the ferromagnetic case we discussed here. The difference is that two types of magnons with opposite polarization coexist in an AFM and they will interplay near the transition to make the sign of the spin current tunable. For an FM, only one type of magnon exists and the sign of the spin current does not change.

{\it Acknowledgments.} HYY acknowledges fruitful discussions with Jiang Xiao and Ming Yan. This project has received funding from the European Research Council (ERC) under the European Unions Horizon 2020 research and innovation programme (grant agreement No. 725509). RD is member of the D-ITP consortium, a program of the Netherlands Organisation for Scientific
Research (NWO) that is funded by the Dutch Ministry of Education, Culture and Science (OCW).

\end{document}